\documentclass[preprintnumbers,twocolumn]{revtex4}
\usepackage{epsfig}

\begin{document}

\title{Orientation-dependent C$_{60}$ electronic structures revealed by photoemission}

\begin{abstract}
We observe, with angle-resolved photoemission, a dramatic change
in the electronic structure of two C$_{60}$ monolayers, deposited
respectively on Ag (111) and (100) substrates, and similarly doped
with potassium to half-filling of the C$_{60}$ lowest unoccupied
molecular orbital. The Fermi surface symmetry, the bandwidth, and
the curvature of the dispersion at $\Gamma$ point are different.
Orientations of the C$_{60}$ molecules on the two substrates are
known to be the main structural difference between the two
monolayers, and we present new band-structure calculations for
some of these orientations. We conclude that orientations play a
key role in the electronic structure of fullerides.
\end{abstract}

\author{V.~Brouet$^{1,2}$, W.L.~Yang$^{1,2}$, X.J.~Zhou$^{2}$, H.J.~Choi$^{3,4}$, S.G.~Louie$^{3,5}$,
M.L.~Cohen$^{3,5}$, A.~Goldoni$^{6}$, F.~Parmigiani$^{6}$,
Z.~Hussain$^{1}$, and Z.X.~Shen$^{2}$}

\affiliation{$^{1}$Advanced Light Source, Lawrence Berkeley
National Laboratory, Berkeley, California 94720\\
$^{2}$ Stanford Synchrotron Radiation Laboratory and Department of
Applied Physics,
Stanford university, Stanford, California 94305\\
$^{3}$ Department of Physics, University of California at
Berkeley,
Berkeley, California 94720\\
$^{4}$ Korea Institute for Advanced Study, 207-43 Cheongryangri
Dongdaemun,
Seoul 130-722, Korea\\
$^{5}$Materials Science Divisions, Lawrence Berkeley
National Laboratory, Berkeley, California 94720\\
$^{6}$ Sincrotone Trieste S.C.p.A., S.S. 14 Km 163.5, in Area
Science Park, 34012 Trieste, Italy }

\date{\today}
\maketitle

In a standard formulation of quantum theory of solids, the
emphasis is on the periodic nature of the lattice structure and
the internal degrees of freedom are usually ignored. As the
frontier of condensed matter physics moves to more complex solids,
such issues become more and more important. Complexity often
arises from situations where interactions with similar energy
scales are competing, and no degrees of freedom can be safely
ignored. Fullerides offer one very interesting example of such a
situation. They are challenging standard approximations in solid
state physics, because electronic correlations in the 3-fold
degenerate band are strong and electrons are coupled to high
frequency phonons \cite{GunnarssonRMP}. The primary reason for
physicists to study them is to understand how these parameters
might lead to new behaviors. However, they are also archetypical
molecular systems, and many degrees of freedom associated with the
C$_{60}$ molecule (e.g. vibrational modes, Jahn-Teller
distortions, orientational order etc.) should be taken into
account, which greatly complicates the analysis. In fact, strong
electronic correlations {\it enhance} the sensitivity to these
local scale structures, because they increase the average time
spent by one electron near a C$_{60}$, so that such a problem is
typically to be expected in a strongly correlated material.

\begin{figure}[b]
\centerline{ \epsfxsize=4.9cm{\epsfbox{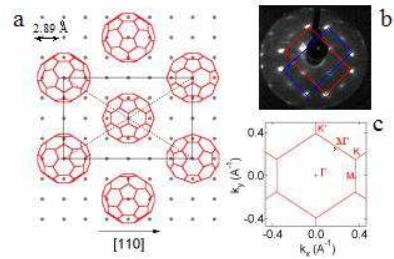}} }
\caption{a) Sketch of the C$_{60}$/Ag(100) structure with respect
to the positions of Ag atoms (grey points). C$_{60}$ are drawn
with the two possible orientations : hexagon-hexagon (6-6) double
bond or pentagon-hexagon (5-6) single bond on top. b) LEED for the C$_{60}$%
/Ag(100) monolayer at 14 eV. Red and blue rectangles define the
reciprocal unit-cell for the two domains. c) First Brillouin Zone
and location of high symmetry points.} \label{Structure}
\end{figure}

We reveal here an extreme sensitivity of the band structure of
C$_{60}$ monolayers to one of this internal degree of freedom,
namely {\it the molecular orientations}. The role of orientations
in the electronic properties of fullerides has often be
questioned. For example, A$_{3}$C$_{60}$ and Na$_{2}$AC$_{60}$
(A=K, Rb), which have similar structures but different
orientational states, are both superconducting but with a
different dependence of the transition temperature on the lattice
parameter \cite{YildirimPRL96,BrownPRB99}.
In the (AC$_{60}$)%
$_{n}$ polymers, a different orientations in C$_{60}$ chains might
control a transition between 1D and 3D electronic structures
\cite{AlloulPRL96}. In TDAE-C$_{60}$, the orientational order can
be changed by the cooling process, which results in different
magnetic ground states \cite{MihailovicScience95}. Nevertheless,
the correlation between electronic properties and orientations has
remained difficult to pinpoint. Recently, we have resolved the
dispersion of a band in a C$_{60}$ monolayer through
angle-resolved photoemission (ARPES) \cite {YangScience03}, which
opens the possibility to monitor directly the changes in band
structure as a function of orientations. A high sensitivity of the
band structure to relative molecular orientations can be expected
because the three degenerate lowest unoccupied molecular orbitals
(LUMOs) that form the conduction band, which are mainly built out
of $p$-orbitals pointing radially at each carbon atom, have high
angular momenta (L=5) \cite{LaouiniPRB95}. We present here an
ARPES study of C$_{60}$ monolayers where structural changes,
including different molecular orientations, are induced by the use
of two different substrates, Ag(111) and Ag(100). We evidence a
complete change of symmetry of the Fermi surface (FS) and of the
band dispersion, and investigate the role of orientations in the
electronic structure with first-principles band-structure
calculations for some of the configurations encountered in these
monolayers.

\medskip

\begin{figure}
\centerline{[t]
\epsfxsize=0.9\columnwidth{\epsfbox{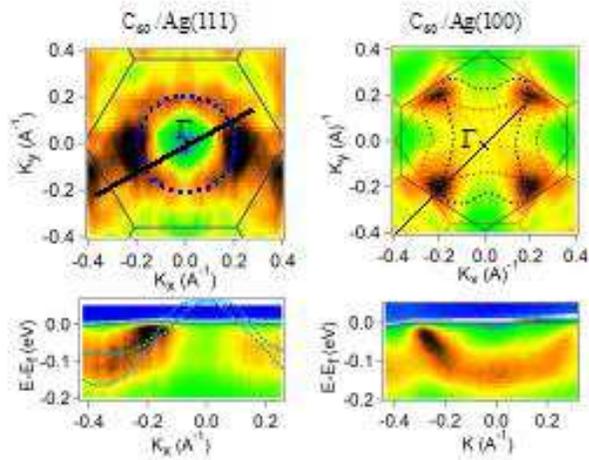}} }
\caption{Comparison of C$_{60}$/Ag(111) and C$_{60}$%
/Ag(100) monolayers near half filling. Top : Map of the spectral
intensity at the Fermi level in the first BZ (see Fig. 3 for color
scale). The dotted lines describe the contour of the FS (for
Ag(100), red and blue refer to the two domains).
Bottom~:~Dispersion along the direction indicated by the thick
black line on the map above.} \label{compare}
\end{figure}

 The growth of C$_{60}$
monolayers on different substrates is very well documented
\cite{ReviewRudolf}. We first deposited a C$_{60}$ multilayer onto
a clean Ag substrate and obtained a monolayer by annealing it at
$\sim $~650 K, and then we doped the layer by potassium (K)
evaporation. The cleanliness of the substrate was checked by the
observation of Ag surface states \cite{AgSurfaceState}, which
disappear after C$_{60}$ deposition. The
structure of the monolayer results from a compromise between the C$_{60}$%
-substrate and C$_{60}$-C$_{60}$ interactions, which are of
similar strength on noble metal surfaces \cite{ReviewRudolf}. The
Ag(111) surface offers the best lattice match with C$_{60}$,
leading to a hexagonal C$_{60}$ overlayer, very similar to a (111)
plane of the bulk compounds. In case of Ag (100), the hexagonal
packing of C$_{60}$ is distorted along one of the two equivalent
directions, as illustrated in Fig. \ref{Structure}a. While this
structure has first been described as c(6*4) \cite{GoldoniPRB}, an
incommensurate structure was proposed more recently
\cite{PaiPRB03}. For our monolayer, the low energy electron
diffraction (LEED) pattern, presented in Fig.~\ref{Structure}b, is
in better agreement with c(6*4), although some distortion from
this model structure might be present. As for the C$_{60}$
orientation on top of the Ag(100) substrate, the scanning
tunneling microscopy (STM) \cite{GrobisPRB} and the X-ray
photoelectron diffraction (XPD) \cite{CepekPRB2001} reveal the
coexistence of two orientations, with either a single (5-6) bond
(between a pentagon and a hexagon) or a double (6-6) bond (between
two hexagons) facing the substrate and being aligned with the
[110] or [1-10] direction
(see examples on Fig.~\ref{Structure}%
a). These orientations contrast with most noble metal (111)
surfaces where a hexagon of C$_{60}$ faces the substrate
\cite{FaselPRL96}, as for Ag(111) \cite{Osterwalder}.

The main ARPES results of the present study are summarized in
Fig.~\ref{compare}, which compares the electronic structure of two
monolayers. In both cases, the number of electrons per C$_{60}$ is
estimated from the integrated area of the LUMO peak to be near 3,
i.e. the band is half filled. All data were collected at the
Advanced Light Source with a 35~eV photon beam in grazing
incidence and polarized nearly perpendicularly to the sample
surface \cite{YangScience03}. In the top of Fig.~\ref{compare}, we
show with dotted lines the contour of the FS that reflects the
different symmetry of the structure of the C$_{60}$ monolayers,
induced by the substrate. The FS is almost circular in the case of
C$_{60}$/Ag(111), while it is rather rectangular for
C$_{60}$/Ag(100). The complete analysis of the FS symmetry was
given in ref. \cite {YangScience03} for Ag(111) and will be given
below for Ag(100). In the bottom of Fig. \ref{compare}, we further
compare the dispersion along high
symmetry lines, which reveals a more unexpected contrast. Most notably, the $%
\Gamma $ point is unoccupied in C$_{60}$/Ag(111),
while it is occupied for at least one of the three LUMO sub-bands in C$_{60}$%
/Ag(100). As the $\Gamma $ point is common to the two
C$_{60}$/Ag(100) domains, this behavior directly establishes a
significant difference in band structures, \textit{regardless of
any further analysis or structural details}. We argue below that
this change is related to the different orientations.

\begin{figure}[t]
\centerline{
\epsfxsize=0.4\textwidth{\epsfbox{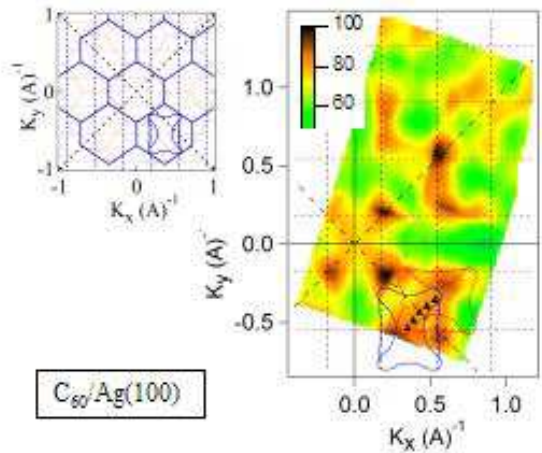}} }
\caption{Fermi surface map for the C$_{60}$/Ag(100) monolayer near
half filling. No symmetry operation were applied to the data.
Inset~: Brillouin zones for the two domains (red and blue) showing
the position of the dotted lines describing the location of the
Fermi Surface (see text).} \label{LargeFS}
\end{figure}

Figure~\ref{LargeFS} presents a larger view of the reciprocal
space in the case of the Ag(100) substrate.  The map was obtained
by integration of the spectral intensity between 0.01~eV and
-0.05~eV from the Fermi level (E$_{F}$). As a result of the
molecular nature of C$_{60}$-based compounds, the photoemission
cross-sections are strongly energy- and angle-dependent \cite
{SchiesslingPRB03}, and particular attention has to be given to
the meaning of the measured intensities. Here, dispersion images
show that each high intensity region of the map corresponds to a
band dispersing towards E$_{F}$, like in Fig.~\ref {compare}. This
rules out a simple modulation of the intensity due to
cross-section or photoelectron diffraction effects
\cite{Photoelectron}. Furthermore, the map presents the
periodicity of the C$_{60}$ reciprocal lattice, whereas such
modulations would be expected over a much larger angular range.
The determination of the FS is complicated in Ag(100) by the
presence of two domains, but their respective contribution can be
distinguished by sampling a large area of momentum space, as in
Fig.~\ref{LargeFS}, because this covers many Brillouin Zones (BZ)
with inequivalent contributions from the two domains. The map is
characterized by a clear symmetry with respect to the diagonals
(black dashed lines), which is actually expected from the
superposition of the two domains (see inset at left of Fig.~\ref
{LargeFS}). Furthermore, the high intensity regions (yellow to
black color) are concentrated along \textit{regularly spaced
vertical and horizontal lines}, shown in Fig.~\ref {LargeFS} as
blue and red dotted lines, respectively. As there is no 4-fold
symmetry for one domain, this regular pattern must originate from
\textit{a well-defined axial symmetry within each domain}, which
will appear as a squaring after superposition. There are two
symmetry axes in the BZ that could play this role, $\Gamma M$ and
$\Gamma K^{\prime }$ (see Fig. 1c), but the spacing between the
dotted lines is only consistent with segments oriented along
$\Gamma K^{\prime }$. This means that the vertical segments of
high spectral intensity arise from the domain drawn in blue, and
the horizontal ones from the one in red. For each domain, the
dotted lines define a ``stripe'', reported on the inset of
Fig.~\ref{LargeFS}, which must contain most of the FS. It can be
worked out that it is the overlap between the bands of the two
domains that blurs the intensity at E$_{F}$ in some regions of the
map (e.g. along the blue line at ky=0).

The clarity of the square pattern implies a simple Fermi surface
geometry, which is in fact surprising when one considers that the
C$_{60}$ LUMOs are triply degenerate, suggesting more than one
piece
 of FS. To create a simple pattern, all LUMO-derived sub-bands must have
similar FS contours mostly following the dotted lines. In fact, we
could only distinguish sub-bands if/when their FS contours
deviate from the dotted lines. There is such a region near $(k_{x}=0.4\,,$ $%
k_{y}=-0.4)$, indicated by black triangles. The dispersion image at k$_{y}$=-0.4~\AA $%
^{-1}$(not shown) reveals two bands crossing the Fermi level with
opposite slopes. This allows to refine the contour of the FS for
these two different sub-bands and the result is shown in
Fig.~\ref{LargeFS}. The larger, more rectangular, contour
corresponds to one (or possibly two) sub-band(s) empty at $\Gamma
$, while $\Gamma $ is filled for the other contour and
corresponding band(s). For clarity, we have reported only the
average contour of these two pieces of Fermi surface in
Fig.~\ref{compare}.

With this knowledge about the FS of C$_{60}$/Ag(100), we can
return in more details to the comparison of Fig.~\ref{compare}.
For C$_{60}$/Ag(111), the monolayer is a single domain and all
directions look roughly the same due to the high hexagonal
symmetry. The comparison with the theoretical band structure
indicates that the observed dispersion corresponds to two
unresolved sub-bands and that the third one remains totally empty
\cite{YangScience03}. For C$_{60}$/Ag(100), we present the
dispersion along the diagonal, where it is the clearest because it
is the same direction, roughly corresponding to $\Gamma M'$, in
both domains. In addition to the difference in curvature at
$\Gamma $, it is interesting to note that the dispersion is
significantly larger for Ag(100) compared to Ag(111), namely 135~meV~$\pm$~$%
15$~meV compared to about 100~meV. Naively, one would expect the
opposite because the distances between C$_{60}$ are larger on
Ag(100) than Ag(111), which should reduce the bandwidth. This
further proves that \textit{the band structure is not simply
``rescaled'' according to the new lattice but much more deeply
modified} and that all parameters must be considered before
comparing two different C$_{60}$ systems.

Two factors come to mind to explain the change in band structure,
either the interaction with the substrate or the orientations of
the C$_{60}$ molecules. It is difficult to estimate possible
contribution from the substrate, but the electronic structure for
the K-doped C$_{60}$/Ag(111) monolayer was found very similar to
that of the bulk \cite{YangScience03}, suggesting only a marginal
influence, as also concluded in Ref.~\cite{HoogenboomPRB98} for
doped monolayers on noble metal surfaces.

\begin{figure}[t]
\centerline{
\epsfxsize=0.45\textwidth{\epsfbox{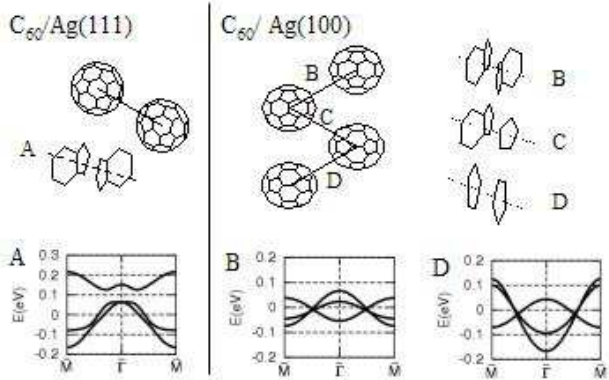}} }
\caption{Top : Sketch of the relative positions of the first C
neighbors for different C$_{60}$ arrangement. The dotted line
joins the center of the two neighboring molecules. Bottom~:~First
principle band structure calculations for some of the
configurations shown above.} \label{neighbors}
\end{figure}

To get a better understanding of the possible role of the
orientations, we now take a closer look at the contact geometry
between two neighboring molecules in the different cases. On
Ag(111), the contact is always through two single bonds, as
sketched in Fig.
\ref{neighbors}A. Note that this ordering of the C%
$_{60}$ molecules is very close to that found in the (111) plane of the disordered $%
fcc$ structure of A$_{3}$C$_{60}$. On Ag(100), more different
contact geometries are encountered, depending on the respective
orientations of two neighboring molecules. A molecule oriented
along a 6-6 bond on top can present either a single bond or a
pentagon to its neighbor. This results in three possible contact
geometries, for this 6-6 orientation only (see
Fig.~\ref{neighbors}) : two single bonds face to face (B), a
single bond towards a pentagon (C), and two pentagons face to face
(D). To investigate the impact of these changes on the electronic
structure, we have calculated the band structure \cite{Cohen} in
cases B and D and found a large difference, as shown on
Fig.~\ref{neighbors}, with bands
reaching much lower energies at $%
\Gamma $ for case D. \textit{This demonstrates the ability of the
orientations to change the electronic structure. }

It is difficult to make a full realistic calculation for Ag(100)
because of the coexistence of different orientations and also of
some uncertainty in the 5-6 orientation (it differs by a few
degrees between STM and XPD \cite{GrobisPRB,CepekPRB2001}).
However, the previous analysis supports the idea that it is the
type of contact geometry that defines the general shape of the
band structure. Indeed, for cases A and B, which are in a similar
(but not identical) configuration, the dispersion is maximum at
$\Gamma $ for at least two bands, in sharp contrast with the very
different configuration represented in D, where two bands show a
deep minimum. STM images show frequent alternation between
different orientations, although in a random way. The structure in
Ag(100) is then likely to be dominated by the ``bond vs polygon''
type of configuration (case C), which is obtained each time two
neighboring molecules have a different orientation. In fact, this
type of configuration is the most favorable energetically because
an electron-rich bond faces an electron-poor polygon. This is what
stabilizes the orientationally ordered $sc $ structure
\cite{LaunoisC60} (that of Na$_{2}$CsC$_{60}$ at low
temperatures), where four different orientations alternate in the
(111) plane to create the ``bond vs polygon'' situation. We
believe that it is these identical contact geometries that allow
the development of a well defined dispersive structure despite the
disorder in orientations. Furthermore, our ARPES result in
C$_{60}$/Ag(100) is in qualitative agreement with the calculation
for the $sc$ structure, based on these contact geometries, where
$\Gamma $ is occupied for two bands \cite{LaouiniPRB95}.

In conclusion, we evidence here for the first time the impact of a
change in C$_{60}$ orientations on the band structure of these
materials. We find that relative orientations can be more
important in defining the band structure than the distances
between molecules. Very early, calculations have suggested that
the band structure could be very sensitive to relative
orientations \cite{SGLouie,GelfandPRL92,LaouiniPRB95}. We present
here new calculations for two different arrangements of the
C$_{60}$ molecules, related to those found in our monolayers,
which reveal differences in band structures in qualitative
agreement with our experimental observation. We show
that the difference in orientations observed in C$_{60}$/Ag(111) and C$_{60}$%
/Ag(100) closely correspond to the different arrangement of the
molecules in a (111) plane of the $fcc$ and $sc$ structures,
respectively. Then, our study gives an experimental basis to the
relevance of orientations in band-structure calculations, not only
for these monolayers, but also to approach real situations in the
bulk, like the difference between A$_{3}$C$_{60}$ and
Na$_{2}$CsC$_{60}$. More generally, this study gives an example of
how the internal structure of the building block of complex
systems can affect their macroscopic properties.

We would like to thank M. Grobis and X. Lu for useful discussion
of their STM data. The SSRL's effort is supported by DOE's Office
of Basic Energy Sciences, Division of Materials Science with
contract DE-FG03-01ER45929-A001. The work at Stanford was
supported by ONR grant N00014-98-1-0195-P0007 and  NSF grant DMR-
0071897. The computational work was supported by NSF Grant No.
DMR00-87088 and BES's Office of the DOE under Contract
DE-AC03-76SF00098. Computational resources have been provided by
NSF at NCSA and by NERSC.

\end{document}